\documentclass[prl,twocolumn,showpacs,superscriptaddress,floatfix]{revtex4}
\usepackage{graphicx}
\usepackage{float}

\begin{document}

\title{Resonant transmission of light through finite chains of subwavelength holes}
\author{J. Bravo-Abad}
\affiliation{\mbox{Departamento de F\'{\i}sica Te\'{o}rica de la
Materia Condensada, Universidad Aut\'onoma de Madrid, E-28049
Madrid, Spain}}
\author{F.J. Garc\'{\i}a-Vidal}
\affiliation{\mbox{Departamento de F\'{\i}sica Te\'{o}rica de la
Materia Condensada, Universidad Aut\'onoma de Madrid, E-28049
Madrid, Spain}}
\author{L. Mart\'{\i}n-Moreno}
\affiliation{\mbox{Departamento de F\'{\i}sica de la Materia Condensada, ICMA-CSIC,
 Universidad de Zaragoza, E-50009 Zaragoza, Spain}}

\begin{abstract}

In this paper we show that the extraordinary optical transmission
phenomenon found before in 2D hole arrays is already present in a
linear chain of subwavelength holes, which can be considered as
the basic geometrical unit showing this property. In order to
study this problem we have developed a new theoretical framework,
able to analyze the optical properties of finite collections of
subwavelength apertures and/or dimples (of any shape and placed in
arbitrary positions) drilled in a metallic film.

\end{abstract}


\maketitle


After the discovery of extraordinary optical transmission (EOT)
through 2D square arrays of subwavelength holes in an optically
thick metallic film \cite{Ebbe98}, several works have appeared in
order to understand the basics of this remarkable phenomenon. From
the theoretical side, the studies can be divided in those
considering the simpler 1D analog of arrays of subwavelength slits
\cite{Schro98,Treacy99,Porto99} and the 2D arrays of holes
\cite{Popov00,Salomon01,LMM01,Sarrazin03}. Several of these works
explained EOT in terms of the existence of surface electromagnetic
(EM) resonances, something pointed out by the original experiments
\cite{Ebbe98} and definitely corroborated by recent experiments
\cite{Barnes04}. However, a question that still remains open is
what is the minimal system that shows EOT, which is interesting
both from the basic point of view and for possible future
applications.

In this letter we move a step forward in this direction and
consider the optical transmission properties of finite chains of
subwavelength holes, a basic structure with less symmetry than the
original 2D array which, up to our knowledge, had not been
considered before. Here we show that EOT is also present in these
1D finite systems. As an important byproduct, we develop a
formalism capable of treating the optical properties of even
thousands of indentations (with any shape and placed arbitrarily)
in metal films, something not possible with the present numerical
methods, which are restricted to just a few of such indentations.

\begin{figure}[h]
\begin{center}
\includegraphics[width=7 cm]{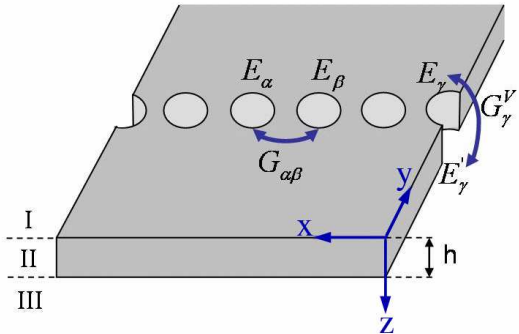}
\end{center}
\vspace*{-1cm} \caption{A chain of circular holes in a metal film
with thickness $h$, with a schematic representation of the terms
appearing in the theoretical formalism presented in the text.}
\end{figure}

Let us first present the formalism, which is a non-trivial
extension of the simpler one developed for sets of 1D indentations
and that was successfully applied \cite{LMM03,FJ03} for the
understanding of enhanced transmission and beaming of light in
{\it single} apertures flanked by periodic corrugations
\cite{Lezec02,Lockyear04}. Here, we analyze the EM transmission
through a planar metal film (with finite thickness $h$ and
infinite in the x-y plane) with a set of indentations at both
input and output interfaces. These indentations may be either
holes or dimples. Furthermore, each one of them may have any
desired shape and may be placed in any position we wish. The only
approximation in the formalism is that the metal is treated as a
perfect conductor ($\epsilon = - \infty$). Then, within this
approximation, the method is virtually exact. We have demonstrated
in previous works that this model captures the basic ingredients
of the enhanced transmission phenomena, being even of
semi-quantitative value in the optical regime for good conductors
like silver or gold \cite{FJ02}. Additionally, results obtained
within the perfect conductor approximation are scalable to
different frequency regimes.

In our method, we assume a rectangular supercell, with lattice
parameters $L_x$ and $L_y$, along the $x$ and $y$ axes,
respectively. This supercell may be real (if we are considering a
bona-fide periodic system) or artificial, if the system is finite.
In this latter case, the limit $L_x, L_y \rightarrow \infty$ must
be taken.

For an incident plane wave with parallel wavevector
$\vec{k}_0$\cite{note_k} and polarization $\sigma_0$, the EM field
at $z=0^-$ (at the metal interface in which radiation is impinging
on) can be written, in terms of the reflection coefficients
$r_{\vec{k}\sigma}$, as
\begin{eqnarray}
|\vec{E}(z=0^-)> &=& |\vec{k}_0\sigma_0 >  + \sum_{\vec{k}\sigma}
r_{\vec{k}\sigma}|\vec{k} \sigma>  \\
-\vec{u}_z \times |\vec{H}(z=0^-)> &=& Y_{\vec{k}_0\sigma_0}
|\vec{k}_0\sigma_0>
 - \sum_{\vec{k}\sigma} r_{\vec{k}\sigma}
Y_{\vec{k}\sigma}|\vec{k}\sigma> \nonumber
\end{eqnarray}
\noindent where we have used the Dirac notation, and expressed the
bi-vectors $\vec{E} = (E_x, E_y)^T$ and  $\vec{H} = (H_x, H_y)^T$
($T$ standing for transposition) in terms of the EM vacuum
eigenmodes, $|\vec{k}\sigma>$. The expressions for these EM vacuum
eigenmodes in real space are: $<\vec{r}|\vec{k} p>= (k_x, k_y)^T
\mathrm{exp}(\imath \vec{k} \vec{r}) / \sqrt{L_x L_y |k|^2} $ and
$<\vec{r}|\vec{k} s>= (-k_y, k_x)^T \mathrm{exp}(\imath \vec{k}
\vec{r})/ \sqrt{L_x L_y |k|^2} $. The electric and magnetic fields
in Eq. (1) are related through the admittances $Y_{\vec{k}s} =
k_z/k_\omega$ and $Y_{\vec{k}p} = k_\omega/k_z$ (for s- and p-
polarization, respectively), where $k_\omega=\omega/c$ ($\omega$
is the frequency and $c$ the speed of light) and
$|k|^2+k_z^2=k_\omega^2$. Notice that, according to Bloch's
theorem, $\vec{k} = \vec{k}_0 + \vec{G}$, $\vec{G}$ being a
(supercell) reciprocal lattice vector.

In the region of transmission, the electric field at $z=h^+$ can
be expressed as a function of the transmission amplitudes
$t_{\vec{k}\sigma}$ as $|\vec{E}(z=h^+)> = \sum_{\vec{k}\sigma}
t_{\vec{k}\sigma}|\vec{k}\sigma>$, from where the magnetic field
can be readily calculated.

The EM fields inside the indentations can be written, in terms of
the expansion coefficients $A_\alpha, B_\alpha$, as:
\begin{eqnarray}
|\vec{E}(z)> &=& \sum_\alpha |\alpha > \left[ A_\alpha e^{\imath
q_{z \alpha} z} + B_\alpha e^{-\imath q_{z \alpha} z } \right] \\
-\vec{u_z} \times |\vec{H}(z)> &=& \sum_\alpha |\alpha > Y_\alpha
\left[ A_\alpha e^{\imath q_{z \alpha} z} - B_\alpha e^{-\imath
q_{z \alpha}z} \right]  \nonumber
\end{eqnarray}
In the previous equations, $\alpha$ runs over all "objects", which
we define as any EM eigenmode considered in the expansion. An
object is, therefore, characterized by the indentation it belongs
to, by its polarization and by the indexes related to the mode
spatial dependence. All that is required to be known are the
electric field bi-vectors $|\alpha>$\cite{normalization} and the
propagation constants $q_{z \alpha}$ associated to the objects, as
the admittance $Y_\alpha= q_{z \alpha}/k_\omega$ for TM modes,
while for TE modes $Y_\alpha= k_\omega/q_{z \alpha}$. For
indentations with such simple cross sections as rectangular or
circular, the required expressions for $|\alpha>$ and $q_{z
\alpha}$ can be found analytically \cite{Morse}; otherwise, they
can be numerically computed \cite{Bell95}. By matching the EM
fields appropriately on all interfaces, we end up with a set of
linear equations for the expansion coefficients. We find it
convenient to define the quantities $E_{\alpha}\equiv
A_{\alpha}+B_{\alpha}$ and $E^{\prime}_{\alpha}\equiv
A_{\alpha}e^{iq_{z \alpha}h}+B_{\alpha}e^{-iq_{z \alpha}h}$, which
are the modal amplitudes of the electric field at the input and
output interfaces of the indentations, respectively. Objects
corresponding to holes are represented by two modal amplitudes,
while objects in dimples require only one (as, in this case,
$A_\alpha$ and $B_\alpha$ are related through the boundary
condition at the dimple closed end). The set $\left\{
E_{\alpha},E^{\prime}_{\alpha} \right\} $ must satisfy:
\begin{eqnarray}
(\epsilon_{\alpha} + G_{\alpha \alpha}) E_{\alpha}+ \sum_{\beta
\ne \alpha } G_{\alpha \beta} E_{\beta}+
G_{\alpha}^{V} E^{\prime}_{\alpha}&=&I_{\alpha} \nonumber \\
(\epsilon_{\gamma} + G_{\gamma \gamma}) E^{\prime}_{\gamma}+
\sum_{\nu \ne \gamma} G_{\gamma \nu} E^{\prime}_{\nu}+
G_{\gamma}^{V} E_{\gamma}&=&0 \label{system_of_eq}
\end{eqnarray}
The different terms in these ``tight-binding'' like equations have
a simple interpretation: $I_{\alpha}\equiv 2 <\vec{k}_0
\sigma_0|\alpha >$ takes into account the direct initial
illumination over object $\alpha$. $\epsilon_{\alpha}$ is related
to the bouncing back and forth of the EM-fields inside object
$\alpha$ and is $\epsilon_{\alpha}= Y_{\alpha}
(1+\Phi_\alpha)/(1-\Phi_\alpha)$ where, for holes,
$\Phi_\alpha=\exp(2iq_{z \alpha}h)$ and the same expression
applies for dimples, but replacing $h$ by $W_\alpha$, the depth of
the dimple. The main difference between holes and dimples is the
presence of $G_{\alpha}^{V}$, which reflects the coupling between
the two sides of the indentation. For a hole
$G_{\alpha}^{V}=2Y_{\alpha}\Phi_\alpha/(1-\Phi_\alpha)$, while for
a dimple $G_{\alpha}^{V}=0$.

The term $G_{\alpha \beta}=\sum_{\vec{k} \sigma} Y_{\vec{k}
\sigma}<\alpha|\vec{k} \sigma ><\vec{k} \sigma|\beta> $ controls
the EM-coupling between indentations. It takes into account that
each point in the object $\beta$ emits EM radiation, which is
"collected" by the object $\alpha$. If the system is periodic,
$G_{\alpha \beta}$ can be calculated through the previous discrete
sum, by including enough diffraction waves. If the considered
supercell is fictitious, the limit $L_x, L_y \rightarrow \infty$
transforms the previous sum into an integral over diffraction
modes. It is then convenient to calculate $G_{\alpha \beta}$
through $G_{\alpha \beta} = <\alpha| \hat{G}| \beta >$. The
integral defining the dyadic
$\hat{G}(\vec{r}_{\parallel},\vec{r}_\parallel^{\
\prime})=<\vec{r}\,|\hat{G}|\vec{r}_\parallel^{\ \prime}>$ can be
evaluated, obtaining
\begin{eqnarray}
\hat{G}_{i j}(\vec{r}_{\parallel},\vec{r}^{\ \prime}_\parallel)&=
&g(d)\, \delta_{i j} + (2\delta_{i j}-1) \, \frac{\partial^2
g(d)}{\partial d_{i} \, \partial d_{j}}
\end{eqnarray}
where $i, j$ can be either $x$ or $y$, $\delta_{i j}$ is
Kronecker's delta, $d \equiv k_\omega
|\vec{r}_{\parallel}-\vec{r}_\parallel^{\ \prime}|$ and $g(d)=-
\imath k_\omega \exp(\imath d)/(2 \pi d)$ is proportional to the
scalar free-space Green function associated to the Helmholtz
equation in 3D.

$\hat{G}$ turns out to be the in-plane components of the EM Green
function dyadic associated to an {\it homogeneous} medium in three
dimensions \cite{Morse}. As a technical note, the calculation of
$G_{\alpha \beta}$ for objects in the {\it same} indentation from
$<\alpha| \hat{G}| \beta >$ suffers from the problems associated
to the divergence of $\hat{G}$ at
$|\vec{r}_{\parallel}-\vec{r}_\parallel^{\ \prime}|\rightarrow 0$
\cite{Yaghjian80,Martin95,Greffet97}, and we evaluate it directly
from its integral over diffraction modes.

Therefore, our method reduces the calculation of EM fields into
finding the EM field distribution right at the indentation
openings, which is extremely efficient when the openings cover a
small fraction of the metal surface. By projecting these fields
into indentation eigenmodes, convergence (as a function of number
of eigenmodes needed) is reached very quickly, especially in the
subwavelength regime. Notice that, although derived in order to
treat finite systems, Eq.(\ref{system_of_eq}) can also be used for
an infinite periodic array of indentations, by imposing Bloch's
theorem on the set $\left\{ E_{\alpha},E^{\prime}_{\alpha}
\right\}$, something we will make use of when analyzing the case
of an infinite linear chain (see below).

Once the self-consistent $\left\{ E_{\alpha},E^{\prime}_{\alpha}
\right\} $ are found, it is straightforward to find all expansion
coefficients, and from them the EM fields in all space. For the
total transmittance through subwavelength holes, we find
$T=\sum_{\alpha}\mathrm{Re}[G_{\alpha}^{V}
E_{\alpha}^{*}E^{\prime}_{\alpha}]/Y_{\vec{k}_0 \sigma_0}$.

We have tested our formalism against published results for two
extreme systems: a single circular hole \cite{Roberts87} and a
square array of square holes \cite{LMM01}. In both cases we
recover the known results, showing that our method is free from
numerical instabilities and that, in finite systems, there are no
spurious effects related to the $L_x, L_y \rightarrow \infty$
limit.

\begin{figure}[h]
\begin{center}
\includegraphics[width=8 cm]{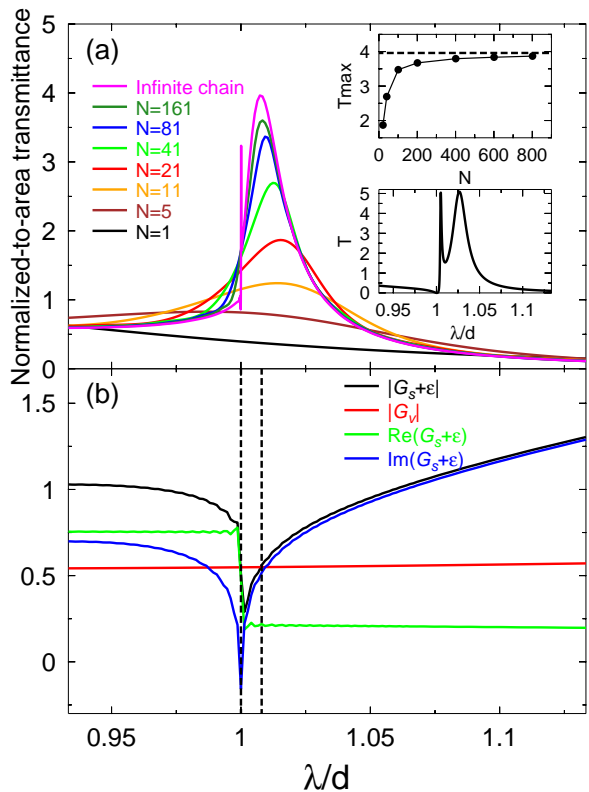}
\end{center}
\vspace*{-0.8cm} \caption{(a) Normalized-to-area transmittance
(see text) versus $\lambda/d$ for a linear array of $N$ holes with
$a/d=0.25$ and metal thickness $h=a$. Top inbox: transmittance
peak value, as a function of $N$, dashed line showing the value
obtained for an infinite chain. Bottom inbox: Normalized-to-area
transmittance versus $\lambda/d$ calculated for an infinite 2D
hole array with the geometrical parameters defining the 1D arrays.
(b) $G_S+\epsilon$ and $G_V$ (see text for the definition of these
magnitudes) as a function of $\lambda/d$.}
\end{figure}

In the rest of this letter we apply this formalism to the study of
linear chains of subwavelength circular holes (see Fig.1). For
proof of principle purposes, we choose holes with radius
$a/d=0.25$ perforated in a metallic film of thickness $h/a=1$, $d$
being the first-neighbor distance between holes. These are typical
geometrical values in experiments in 2D hole arrays. Fig.2a shows
the evolution of the transmittance versus wavelength as a function
of the number of holes, $N$ \cite{two_modes}. The incident plane
wave impinges normally, and is polarized with the E-field pointing
along the direction of the chain. For the other polarization, the
boost in transmittance is negligible. In Fig.2a, the total
transmission is normalized to the hole area and then divided by
the corresponding $N$. The most interesting feature of these
spectra is that, as $N$ is increased, a transmittance peak emerges
at $\lambda$ close to $d$, showing that enhanced transmission is
also present in linear chains of subwavelength holes. The
transmittance peak value, $T_{max}$, grows almost linearly with
$N$ (for small $N$) eventually reaching saturation (see top inbox
of Fig.2a). In order to gain physical insight into the origin of
this transmission resonance, it is interesting to analyze a
simplified model for the infinitely long chain. In this model,
only one mode per hole is considered: the least evanescent mode
with the electric field pointing mainly along the chain axis.
Therefore, for all wavelengths, each hole behaves as a small
dipole. In this geometry, Bloch's theorem implies $E_{\alpha} = E
\,  \mathrm{exp}(\imath k_x \alpha d)$, and $E_{\alpha}^{\prime} =
E^{\prime}\,  \mathrm{exp}(\imath k_x \alpha d)$, which renders
the system of equations (3) easily solvable. For the case we are
considering (normal incidence, $k_x=0$) this procedure yields,
\begin{equation}
[(\epsilon + G_{S})^2-G_{V}^{2}] E=I(\epsilon+G_{S})
\end{equation}
\noindent where $G_{S}=G_{\alpha \alpha}+\sum_{\beta \ne
\alpha}G_{\alpha \beta}$, $G_V=G_{\alpha}^{V}$, and
$\epsilon=\epsilon_{\alpha}$. Fig. 2b shows that the spectral
location of the transmittance peaks for the infinite chain
coincide with the cuts between $|\epsilon + G_{S}|$ and $|G_{V}|$,
implying that the origin of EOT relies on a resonant denominator.
Therefore, EOT is associated, as in the case of 2D hole arrays, to
the excitation of coupled surface EM resonances, which radiate
into vacuum as they propagate along the surface. This is yet
another instance of surface EM modes (in this case a leaky mode)
appearing in a perfect conductor due to the presence of an array
of indentations\cite{Pendry04}. The results for finite chains (top
inbox of Fig.2a) show that this EM resonance is characterized by a
typical length, $L_D$ (for the case considered $L_D\approx 80 d$)
. When the size of the finite chain is smaller than $L_D$, the
resonance is not fully developed and the transmittance is smaller
than the one obtained for an infinite chain. However, for large
enough $N$, the system can effectively be considered as infinite
and its associated transmittance peak approaches the asymptotic
value (with a $1/N$ contribution coming from holes located at a
distance $\approx L_D/2$ from the chain ends). We have found that
$L_D$ is completely governed by the geometrical parameter $a/d$,
increasing as $a/d$ decreases.

\begin{figure}
\begin{center}
\includegraphics[width=8 cm]{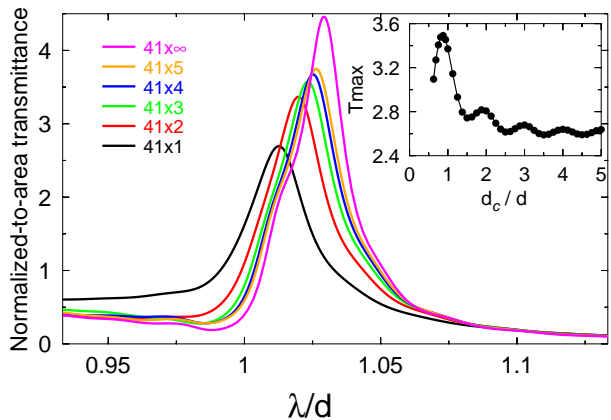}
\end{center}
\vspace*{-0.8cm} \caption{Normalized-to-area transmittance versus
$\lambda/d$ for several stripes formed by $M$ linear chains of
$41$ holes with $a/d=0.25$ and $h/a=1$. $M$ varies from $1$ to
$5$. Also the limiting case of $M=\infty$ is presented. Inset:
Transmittance peak value for the case of two linear chains (each
one of them with the same parameters as before), as a function of
their distance, $d_c$.}
\end{figure}

It is also interesting to compare the peak value calculated for an
infinite linear array ($T_{max}^{1D} \approx 4$, see Fig.2a) with
the one obtained for an infinite 2D hole array with the same
geometrical parameters ($T_{max}^{2D} \approx 5$, see bottom inbox
of Fig.2a). This is, when going from a linear chain to a 2D hole
array, the transmittance per hole is only increased by $25 \%$.
This means that the basic unit of the extraordinary transmission
phenomenon observed in 2D hole arrays is the linear chain of
holes, and that the 2D hole array can be considered as an array of
weakly coupled 1D arrays. Although in this paper we only show
results for normal incident radiation, we have checked that this
conclusion remains valid provided the in-plane component of the
incident E-field points along the direction of the chain.

In order to study how the EOT in 2D hole arrays develops from the
EOT in linear chains, we have calculated the transmittance of a
collection of finite linear chains separated by a distance $d$.
Figure 3 renders the normalized-to-area transmittance versus
wavelength, for different stripes formed by several (ranging from
$1$ to $\infty$) chains of $41$ holes. As shown in the figure, the
stronger effect appears when going from $41\times 1$ to $41 \times
3$, suggesting that the EM coupling between chains of
subwavelength holes is very short-ranged.  The short-range nature
of the inter-chain interaction is more clearly seen in the inset
to Fig. 3, which renders the transmission peak value, $T_{max}$,
through two finite linear chains as a function of the distance
between them, $d_c$. As this inset shows, for this set of
geometrical parameters, the chains are already practically
uncoupled when $d_c=3d$, the maximum coupling being for $d_c
\approx d$.

When linear chains are added up to the structure, the
transmittance peak shifts to longer wavelengths and its maximum
value increases. The increase in transmittance is due to the fact
that each chain takes advantage from the re-illumination coming
from other chains. The peak redshift is related to the
corresponding decrease of frequency of the stripe surface EM mode,
due to a reduction of its lateral confinement. More precisely, in
this case the stripe surface mode that couples to the incident
plane wave with $\vec{k}=0$ is essentially the sum, with equal
phases, of the single chain leaky modes. In infinite 2D hole
arrays, two EOT peaks originate from the resonant coupling (via
the holes) of the surface EM modes, with a narrow transmission
peak appearing close to $\lambda \approx d$ (see bottom inbox of
Fig.2). Notice that, in stripes of chains of finite length only
one peak is clearly resolved, the second expected peak appearing
as a shoulder in the transmission curve (see Fig. 3). This
suggests that finite size effects may prevent the development of
the narrowest peaks. Interestingly, the addition of linear chains
also provokes the birth of a minimum in the transmittance
spectrum, appearing at a wavelength slightly smaller than $d$.
Eventually, in 2D hole arrays, this minimum leads to a ``Wood's
anomaly'', appearing just when a propagating Bragg diffracted wave
becomes evanescent. In this case the reciprocal lattice vectors
involved are $(\pm 1,0) 2 \pi / d$, and Wood's anomaly appears due
to a divergence in the EM density of states corresponding to those
wavevectors. Notice that, in a linear chain, as the diffracted
field contains a continuum of $k_y$ components, the divergence in
the density of states is smeared out; correspondingly no Wood's
anomalies appear in this case.

To summarize, we have found that extraordinary optical
transmission phenomenon is already present in a single finite
chain of subwavelength holes in metallic films. For a chain, the
transmittance per hole is comparable to that found in 2D arrays;
therefore, the single chain can be considered as the basic entity
of EOT and then 2D arrays can be seen as a collection of weakly EM
coupled chains. As a byproduct we have developed a new theoretical
framework which is able to treat the optical properties of even
thousands of indentations (holes or dimples) placed arbitrarily in
a metallic film. This numerical tool will surely help in the
design of new arrangements of subwavelength objects pursuing
specific optical functionalities.

Financial support by the Spanish MCyT under grant BES-2003-0374
and contracts MAT2002-01534 and MAT2002-00139, and the EC under
project FP6-NMP4-CT-2003-505699 is gratefully acknowledged.


\begin{thebibliography}{}
\bibitem{note_k} For wavevectors, we use the notation
$\vec{k}=(k_x,k_y)$.
\bibitem{normalization} We normalise the modal bi-vectors so that
$\left<\vec{k} \sigma|\vec{k} \sigma\right>=1 $ and
$\left<\alpha|\alpha\right>=1 $.
\bibitem{two_modes} Considering just the two least decaying modes in each hole
is enough to reach very accurate results for this geometrical
setup.
\end{thebibliography}
\end{document}